\documentclass[twocolumn,floats,aps,showpacs]{revtex4b4}
\usepackage{times,amsmath,graphics,psfrag,latexsym}

\begin{document}
\title{Smooth Vortex Precession in Superfluid ${}^4$He}
\author{L. Hough, L.A.K. Donev, and R.J. Zieve}
\affiliation{Physics Department, University of California at Davis}
\begin{abstract}
We have measured a precessing superfluid vortex line, stretched from
a wire to the wall of a cylindrical cell.  By contrast to previous
experiments with a similar geometry, the motion along the wall is smooth.
The key difference is probably that our wire is substantially off center.
We verify several numerical predictions about the motion, including an
asymmetry in the precession signature, the behavior of pinning events,
and the temperature dependence of the precession.
\end{abstract}
\pacs{47.32.Cc, 47.37.+q, 67.40.Vs}
\maketitle

\section{Introduction}

Vortices appear in systems from superconductors and superfluids to weather
patterns and blood flow \cite{gen}.  They underlie complex fluid flows, and 
motivate computational techniques \cite{vortmeth1, vortmeth2}.  Measurements of the 
behavior of one or a few vortices provide test cases for computations and
starting points for  understanding more complicated vortex systems.  Indeed,
even small numbers of vortices can produce interesting dynamics, such as the
hopscotch motion of two coaxial vortex rings \cite{2rings}.  Yet in classical
systems, even {\em defining} a single vortex can be difficult \cite{gen2}.  For
example, in a classical fluid the vorticity fields of two nearby vortices may
overlap and resemble a single vortex.  

By contrast, a superfluid vortex is well-defined, since circulation is
quantized and non-zero vorticity occurs only within the vortex core.  In
superfluid ${}^4$He, the core size is only a few angstroms, 
and vortices closely approximate
the ideal slender vortices of theoretical treatments.  Although in principle
superfluid experiments can verify theoretical and computational models of
vortex dynamics, few measurements can probe individual vortex behavior.  Two
exceptions are the observation of a precessing vortex in a Bose-Einstein
condensate \cite{BEC} and diffraction experiments on superconducting vortices
\cite{Finnemore}.  However, these techniques have limitations for dynamical
studies, the former because of the condensate lifetime and the latter because
of the time for each measurement.  Here we discuss a different experiment,
which detects single vortex motion in superfluid helium. 

Our apparatus consists of a straight wire, stretched along a cylinder.
A current pulse in a perpendicular magnetic field makes the wire vibrate.
The subsequent motion perpendicular to the field
is detected as a voltage across the wire.  In the presence of fluid
circulating around the wire, the normal modes of
the wire's vibration are split, with the beat frequency revealing the
circulation around the wire.  This technique originated nearly forty
years ago in early measurements of quantized circulation in superfluid
${}^4$He \cite{Vinen, WZ}, and was later used to measure circulation in superfluid
${}^3$He \cite{3HeQC}.

A question that persisted through two decades of vibrating wire measurements is
why intermediate circulation values are frequently observed.  The expected
quantum levels are clearly more stable, but other values occur as well.  One
suggestion was that a vortex might run along the wire for only part of the
wire's length, then detaching and becoming a free vortex in the fluid,
terminating on the cell wall.  This results in a partial effect on the
vibration frequencies, as would fractional circulation along the entire length
of the wire.  Recently this idea was verified, originally in superfluid
${}^3$He \cite{helicopter, 3Hepaper}.  The free end of a partially trapped
vortex precesses about the wire, driven by the flow field of the trapped
portion.  If the wire is displaced from the axis of the cell, the precession
appears in the beat frequency as oscillations.

A similar effect was observed in ${}^4$He, with the same cell used for the
${}^3$He measurements, although noise in the attachment point motion usually
hid any oscillations present \cite{4Hepaper}.  Even the cleanest ${}^4$He
signal was far more irregular than a typical ${}^3$He  signature.  Furthermore,
the earlier generations of vibrating wire measurements in ${}^4$He found no
recognizable precession signature.  One possible explanation of the difference
in signals is the much smaller vortex core radius: 1.3 \AA \ in ${}^4$He as
opposed to 1000 \AA \ in ${}^3$He.  The smaller core makes the ${}^4$He
vortex more sensitive to roughness on the surface of the cell and the wire. 
Given the difficulty of experiments on rotating superfluid ${}^3$He,
particularly vibrating wire experiments that must be carried out well below
the 2 mK transition temperature,
and the appeal of further measurements on a single vortex line undergoing
a fundamentally three dimensional motion, we
returned to ${}^4$He to attempt to resolve a smoother signature there.

Here we report smooth vortex precession in ${}^4$He.  In the process of this
work, we have verified several numerical predictions about the details of the
motion \cite{Schwarzheli}, which were inspired by the original ${}^3$He work.
 
\section{Experimental background}
Our experimental setup is similar to that used in previous generations of
vibrating wire experiments \cite{Vinen,WZ,4Hepaper}. A brass cylinder of height
50 mm and inner radius about 1.5 mm is filled with  helium.  A NbTi
wire a few microns in radius is stretched parallel to the cylinder's axis, 
entering the cell through
Stycast 1266 caps with small holes displaced 0.3 to 0.5 mm from the axis.  The 
wire is glued into the caps with a small weight hanging from it to keep it
under tension.

In principle the wire should have doubly degenerate normal modes at its
fundamental frequency.  However, as explained in detail in Ref. \cite{WZ}, the
modes are actually linear and perpendicular.  They differ in frequency by 
$\Delta\omega_0$, probably because of imperfections in the wire's cross-section
or mounting.  If the wire serves as the core of a superfluid vortex with
circulation $\kappa$, the normal modes become ellipses with major axes aligned
with the original linear modes.  The frequency splitting increases to 
\begin{equation}
\Delta\omega = \sqrt{(\Delta\omega_0)^2+(\frac{\rho_s\kappa}{\mu})^2},
\label{e:quad}
\end{equation}
where $\rho_s$ is the
superfluid density and $\mu$ is the mass per length of the wire, adjusted for
the fluid displaced during vibration.  The lowest non-zero circulation 
is $\kappa=\frac{h}{m}$, with $m$ the mass of a ${}^4$He atom.  With 
$\rho_s=0.145$ g/cm$^3$ and $\mu\approx 15\times 10^{-6}$ g/cm, the frequency
splitting from this circulation is $\frac{\rho_s\kappa}{\mu}=9.6$ rad/s.

The magnetic field, about 200 Oersted, comes from two saddle coil magnets.   
Changing their relative current rotates the field through the plane perpendicular to
the wire.  For detection purposes, a key point is that in general the wire does
not move in a plane perpendicular to the magnetic field.  Since the excitation
method uses current in the wire and the resulting Lorentz force, 
the wire always {\em starts} moving perpendicular to the field;
but vibration in this plane usually is not a normal mode.  Instead the wire
traces out a elaborate path, with its velocity  at times nearly or entirely 
perpendicular to the magnetic field.  The induced emf shows beats in the
oscillation envelope at frequency $\Delta\omega$. Orienting the field halfway
between the two linear normal modes excites  the modes equally and leads to
complete beats, with or without circulation around the wire.  

The voltage signal goes
to a 100:1 PAR 1900 transformer at room temperature, then to a Stanford
Research Instruments SR560 pre-amplifier and  SR830 lock-in amplifier.  The
oscillation magnitude is read to the computer through a 16-bit A/D board.  In
real time, we fit the envelope to an exponentially damped sine wave with four
adjustable parameters (amplitude, phase, frequency, and damping).  We also save
the original digitized decays for possible re-analysis later.  The accuracy in
ascertaining the beat frequency varies from wire to wire.  It depends heavily,
for example, on the beat frequency itself.  Typically we can extract the beat
frequency to 0.1\% below 400 mK.  The uncertainty arises primarily from
mechanical vibrations which shake the wire, rather than from electronic noise.

A partially trapped vortex has less effect on the wire's beat frequency.  The
beat frequency in this case depends on what part of the wire is covered with
the vortex. Since the middle has a larger vibration amplitude than the ends, it
contributes more heavily to the final beat frequency.  If there is one quantum
of circulation on the wire below an attachment point $z$, and zero circulation
above $z$, then Equation \ref{e:quad} still holds, with $\kappa$
replaced by an effective circulation $<\kappa>$ 
defined by \cite{WZ}
\begin{equation}
<\kappa>=\frac{1}{2L}\int_0^z \frac{h}{m} \sin^2\frac{\pi}{L}z^\prime dz^\prime=
\frac{h}{m}(\frac{z}{L}-\frac{1}{2\pi}\sin\frac{2\pi z}{L}).
\label{e:position}
\end{equation}
Here $L$ is the length of the wire, and we assume that the lowest normal modes
are sinusoidal in the $z$ direction.

\begin{figure}[b]
\begin{center}
\scalebox{.5}{\includegraphics{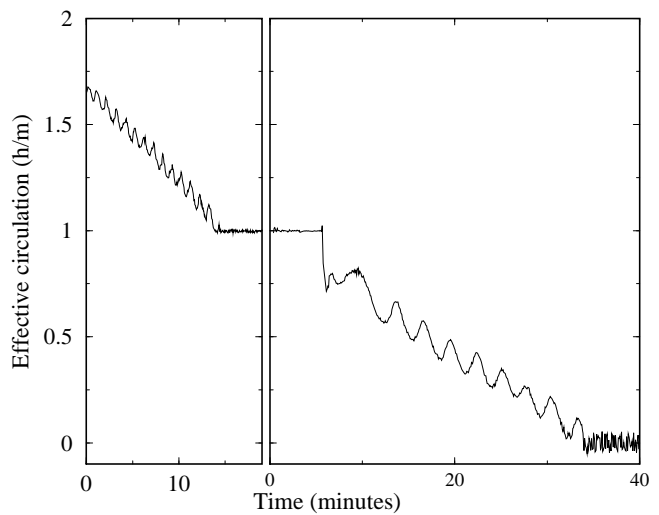}}
\caption{\small Unwinding of vortex from wire.  The two parts of the figure
have the same horizontal scale.}
\label{f:2to0}
\end{center}
\end{figure}

The conversion from the total beat frequency to the contribution from
circulation, Equation \ref{e:quad}, is insensitive at values of 
$\Delta\omega$ near $\Delta\omega_o$, where the wire's imperfections overwhelm
the effect of circulation. The conversion from circulation to attachment point
location, Equation \ref{e:position}, has low sensitivity near both $<\kappa>=0$
and $<\kappa>=\frac{h}{m}$. As a result, our measurements are mainly of the
vortex moving in the middle of the cell, where our sensitivity lets us locate
the attachment point to better than 50 $\mu$m.

\begin{table*}[htb]
\caption{Estimated dimensions for the five wires used here.}
\label{t:wiredata}
\newcommand{\m}{\hphantom{$-$}}
\newcommand{\cc}[1]{\multicolumn{1}{c}{#1}}
\renewcommand{\tabcolsep}{1pc} 
\renewcommand{\arraystretch}{1.2} 
\begin{tabular}{@{}lccccccccc}
\hline
& $R$ (mm) & $R_W$ ($\mu$m) & $\omega$ (rad/s) & 
$\Delta\omega_o$ (rad/s)& Smallest $T$ (s) & Predicted $T$ (s)\\
\hline
A & 1.4 & 5.1 & 6666 & 25.5 & 187 & 138 \\
B & 1.4 & 5.1 & 4115  & 3.02  & --- & 138 \\
C & 1.4 & 8.9 & 2262  & 6.53 & 147 & 154 \\
D & 1.53 & 8.8 & 1929  & 10.3 & 171 & 180 \\
E & 1.53 & 8.6 & 1030  & 33.8 & 194 & 179 \\
\hline
\end{tabular}\\[2pt]
\end{table*}

As mentioned previously, when a vortex is partially attached to the wire, the
free end precesses about the wire.  If the wire is displaced from the axis of
the cylinder, the length of the free vortex changes as it moves.  This would 
alter the energy in the flow field unless compensated by motion of the
attachment point.  We detect the attachment point motion as oscillations in the
beat frequency of the wire.

Figure \ref{f:2to0} shows the oscillations characteristic of the unwinding
process.  The two portions of the figure show data taken at different times
but with the same cell.
Each point represents a separate plucking of the wire.  The
oscillations correspond to the motion of the free vortex around the cell, and
the gradual decay of the circulation happens because the
moving vortex dissipates energy.  Figure \ref{f:2to0} also shows the 
qualitative distinction
between the expected quantized circulation levels, which we will refer to by
quantum number ($n=0$, $n=1$, etc.) and intermediate levels, both at $n>1$ and
$n<1$.  The quantized levels  are far more stable and exhibit none of the
oscillations due to precession.  The increased noise near $n=0$ is inherent in
the analysis equations, as mentioned above.  Finally, the figure shows the
different precession periods for decays at $n>1$ and $n<1$, roughly
one minutes and three minute, respectively.  In the former case
the circulation around the wire is $2\frac hm$ near one end and $\frac hm$ near the
other, with a singly quantized vortex emerging from the wire at the spot where
the circulation changes.  The flow field is that of three
half-vortices, rather than one, so the vortex precesses three times as fast.

We discuss measurements on five wires, which are described in Table
\ref{t:wiredata}.  Each wire is a single strand cut from multistranded NbTi
wire.  By weighing several strands, and using 6.0 g/cm$^3$ as the density
of NbTi, we find that the average mass per length
corresponds to a radius of 5.1 $\mu$m for wires such as A and B, and 8.7 $\mu$m
for wires such as C, D, and E.  For the wires that trap a stable $n=1$ vortex,
we can take the observed stable beat frequencies as $\Delta\omega$ and
$\Delta\omega_o$ in Equation \ref{e:quad} and find the mass per length of the
wire.  The corresponding radius values, shown in Table \ref{t:wiredata}, are
close to the values found from the weights. For wires A and B, no stable $n=1$
state was seen, and Table \ref{t:wiredata} shows the radius calculated from the
typical wire weight.  As an additional check, the room temperature resistance
of the wires should be proportional to cross-sectional area.  The radius values
in Table \ref{t:wiredata} are indeed consistent with the wires' resistances. 
As discussed below, the amplitude and shape of the oscillation signals depend
on the displacement from the cylinder axis. Our observations are consistent
with the displacements measured during assembly of the cells.

\begin{figure}[b]
\begin{center}
\scalebox{0.5}{\includegraphics{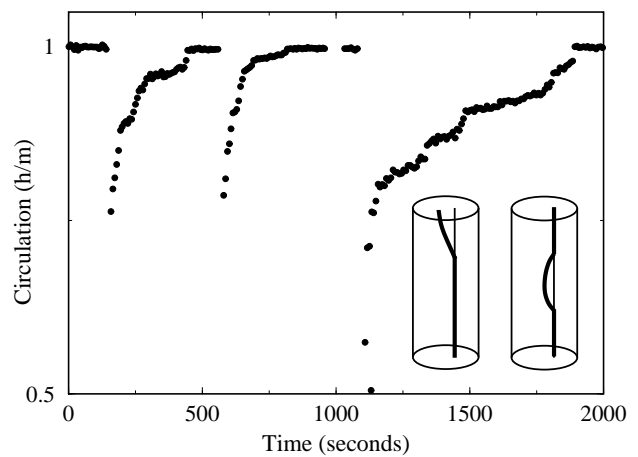}}
\caption{\small Characteristic signal for vortex partly dislodging and
returning to wire.  The insets show two possibilities for the vortex
configuration corresponding to the intermediate circulation values.}
\label{f:windon}
\end{center}
\end{figure}

The measurements are performed on a pumped ${}^3$He cryostat with a
minimum temperature of 250 milliKelvin.  The cryostat has an external
${}^4$He pumping line, which passes through a rotating vacuum seal.
For the experiments discussed here, all measurements are made with the
cryostat stationary.  We fill the cell at 4K, then cool into the superfluid.
Once cold, we rotate by hand, typically 10 revolutions in one minute, to 
create circulation.  During rotation our electrical wires are disconnected.
When we restart the measurements, we usually find an unstable state, which
can have circulation either larger or smaller than the $n=1$ state.  If the
circulation settles to a stable $n=1$ state,
we dislodge the vortex by moving the cryostat rapidly about 60 degrees, in
the direction opposite to the previous rotation direction, and then slowly
returning it to its original position.  This minimally controlled shaking
generally results in one of two vortex signatures.  Sometimes an
abrupt drop in trapped circulation is followed by oscillations indicating
vortex precession.  At other times, the circulation drops, but then
increases back to $n=1$ in the characteristic, slightly irregular
way illustrated several times in Figure \ref{f:windon}.  One possible
explanation is that a portion of the vortex is dislodged from the wire,
but does not reach the cell wall.  The detached portion then
precesses through the cell, spiraling inward until it
is recaptured by the wire.  We have not yet identified whether the
detaching occurs near the center or end of the wire; schematics of these
two possibilities are shown in Figure \ref{f:windon}.

\section{Computational background}
To complement our experimental work, we do simulations of superfluid vortex
dynamics.  Our code is similar to that of Schwarz \cite{Schwarzheli, Schwarz},
and is described elsewhere \cite{lukesim}.  Here we briefly summarize its
workings.  The equations of motion for the vortex are based on the
incompressible Euler equations, which would require the vortex core to move at
exactly the local superfluid velocity.  A friction term $-\alpha {\bf s^\prime}
\times {\bf v_s}$ is added which gives the vortex motion an additional
component perpendicular to the superfluid velocity.  Here the friction
coefficient $\alpha$ determines the strength of the friction interaction; ${\bf
s^\prime}$ is the tangent to the vortex line, directed so that the circulation
is given from ${\bf s^\prime}$ by the right-hand rule; and ${\bf v_s}$ is the
superfluid velocity.  We calculate the fluid velocity from the instantaneous
core locations and boundary conditions imposed by the cylindrical container. 
The vortex cores have a local contribution proportional to the vortex
curvature, and a non-local contribution from distant segments which obeys the
Biot-Savart law.  For the present geometry, we neglect the latter except for
the portion of the vortex trapped along the wire, which we treat as a
half-infinite straight  vortex.  We meet the boundary conditions approximately
using an image vortex: a second half-infinite vortex defined from the trapped
vortex by inversion in the cylinder.  Adding the image under inversion of the
free vortex segment does not significantly change the results. 

We use dimensions similar to those of the actual experiment:  a cell of
radius 1.5 mm, and a wire of radius 6 $\mu$m.  Near the wire the point
spacing along the vortex is comparable to the wire radius, and it gradually
increases with increasing distance  from the wire.  

\section{Smooth precession}
In the absence of dissipation, the precession period for a single partially
attached vortex is given by
\begin{equation}
T=\frac{4\pi^2(R^2-R_W^2)}{\kappa\ln\frac{R}{R_W}}.
\end{equation}
As shown in Table \ref{t:wiredata}, the predicted and actual values for our
cells are in reasonable agreement, with typical precession periods slightly
under three minutes.  The discrepancies are larger than in ${}^3$He 
\cite{helicopter, 3Hepaper} or in lower-temperature ${}^4$He \cite{4Hepaper}
measurements.  A major reason for this is that in our temperature range the
precession period varies substantially with temperature.  We discuss this
behavior below.  

We first address the smooth nature of the motion, not previously found in
${}^4$He.  Four of the five wires show smooth precession signatures.  With the
fifth, wire B, circulation decays rapidly with no clean precession. These are
the {\em only} wires we have measured, so the contrast with the noisy
intermediate-circulation data from the previous experiments is striking.  The
one deliberate difference in our design is that, instead of trying to center
the wire in the cylinder, we intentionally move it off the axis to increase the
amplitude of the precession signal.  We suggest that a displaced wire
also leads to more regular vortex precession.

Hints that wire displacement can influence the smoothness of the precession
first appear in Schwarz's simulations of a vortex precessing in a cylinder
\cite{Schwarzheli}. His calculations incorporate pinning, but without any
geometrical details of the pin sites; rather, one end of the vortex is simply
fixed in position.  The local fluid velocity bends the vortex towards the
wall.   For depinning, Schwarz uses a critical angle model:   when the angle
between the vortex and the wall falls below a critical value, the fixed end of
the vortex becomes free, and is once again allowed to move along the wall. If
the critical angle is too small, the vortex never reaches it and remains
pinned. To simulate vortex precession with wall roughness, the chosen angle
must be large enough that the vortex never becomes permanently pinned. Schwarz
mentions that an off-center wire allows a smaller critical angle than an
on-center wire \cite{Schwarzheli}.  This suggests that a vortex running from an
off-center wire to the wall is better able to free itself from pin sites and
continue its motion.  Our present work gives some experimental confirmation  of
the critical angle model.  The reduced pinning from putting the wire
deliberately off center changes the signature from noisy to smooth.

\begin{figure}[htb]
\begin{center}
\scalebox{.99}{\includegraphics{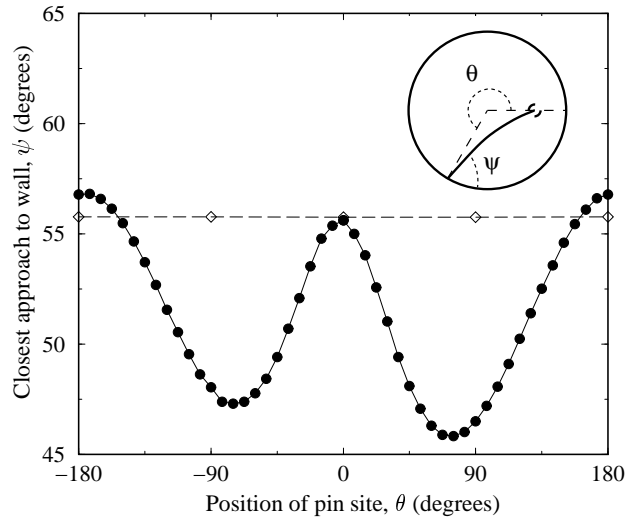}}
\caption{\small Minimum angle from wall that occurs as vortex settles
into pin site.  The calculations use no friction ($\alpha=0$). 
$\diamond$: wire on cylinder axis.  
$\bullet$: wire displaced 0.4 mm from cylinder axis.  The inset shows the
geometry.  The black dot represents the wire, which has counterclockwise
circulation on the portion away from the viewer.  The solid arc is the 
free segment of the vortex.  The angle $\psi$ may not lie in the plane
of the page.}
\label{f:critangle}
\end{center}
\end{figure}

Our own simulations on pinning in a cylinder provide more details.   We begin
by calculating the trajectory of a precessing vortex.  Initially waves
propagate along the vortex as it settles into the configuration appropriate 
for steady precession.  Once these oscillations in the vortex shape have
decayed away, we abruptly cease the motion of the endpoint on the wall.
The vortex then again oscillates, settling gradually into a permanent
position.  We find the minimum angle with the wall
achieved during this process, shown in Figure \ref{f:critangle} for an
on-center wire and one displaced 0.4 mm.  The point spacing near the wall is 53
$\mu$m for these calculations, corresponding to a bump size of 23 $\mu$m
\cite{lukesim}. As the wire becomes farther off center, the minimum angle
decreases substantially at most locations in the cell, but increases slightly
where the vortex is longest.  Our interpretation is that a vortex coming from
an off-center wire frees itself more easily from a pin site, for most locations
of the pin site.  However, it appears that near $180^\circ$ the off-center wire
actually makes depinning more difficult. Experimentally the smooth precession 
does not cease in this portion of the cell, so some further explanation is
required. One explanation is to model the wall roughness as bumps of
different sizes, each with its own critical angle.  Larger bumps, which pin
vortices more easily, correspond to smaller critical angles.  As long
as the largest bumps lie outside the narrow region near $180^\circ$ where the
pinning is easier for the displaced wire than for the on-center wire, moving
the wire off center may still reduce pinning.  Another possibility is that
encountering pin sites during precession alters the vortex shape, 
presumably bending the vortex line closer to the cell wall.  Changing the
configuration at the onset of a pinning event affects the minimum angle
achieved, perhaps decreasing the off-center wire's minimum angle below that of
the on-axis wire everywhere in the cell.

Several factors determine the angle the vortex attains after pinning.  The flow
velocity depends on the displacement of the wire from the cell axis, and also
on the location of the free vortex in the cell.  The velocity is faster where
the wire approaches the cell wall more closely. In addition, if the wire is off
center, a straight line from the wire to the cell wall is generally not
perpendicular to the wall.  This defines a natural angle for a vortex at any
spot in the container.  Although the vortex may curve, meeting the wall at a
different angle, this simple geometrical consideration is a major factor in
determining the final angle. 
Finally, the asymmetry visible between the two sides of the cell comes from the
direction of the fluid flow relative to the natural angle of the vortex.

Apart from the displaced wire, nothing about our setup differs substantially
from two previous sets of experiments at temperatures
below 400 mK \cite{4Hepaper, Karn}.  Our cylindrical cells have comparable
dimensions to those in the earlier work.  They are drilled or reamed,
with no special polishing or other treatments, so the cell walls are probably
no smoother than in the earlier experiments.  Although our wires are the
smallest used to date, our larger wires are within a factor of 2 of nearly all
those previously reported.   We believe the wire size improves the precession
signatures only in a minor and indirect way. Most of our measurements are on
vortex precession between $n=1$ and $n=0$, although we also observe precession
between $n=2$ and $n=1$, as shown in Figure \ref{f:2to0}.  In previous
experiments slow transitions were usually found only between the higher
circulation levels, with $n=1$ so stable that only steadily rotating the
cryostat in the opposite sense to the vortex would dislodge it.  From Figure
\ref{f:2to0}, the $n=2$ to $n=1$ precession may be slightly noisier than that
from $n=1$ to $n=0$, but it is clearly visibly and far cleaner than previous
${}^4$He precession data \cite{4Hepaper}.  The lower-circulation motion may be
more regular because less vorticity remains in the cell.  If so our smaller
wires smooth the precession signature by favoring precession at $n<1$, but are
not a crucial feature of the current experiment.

\section{Pinning}
\begin{figure}[bht]
\begin{center}
\scalebox{.5}{\includegraphics{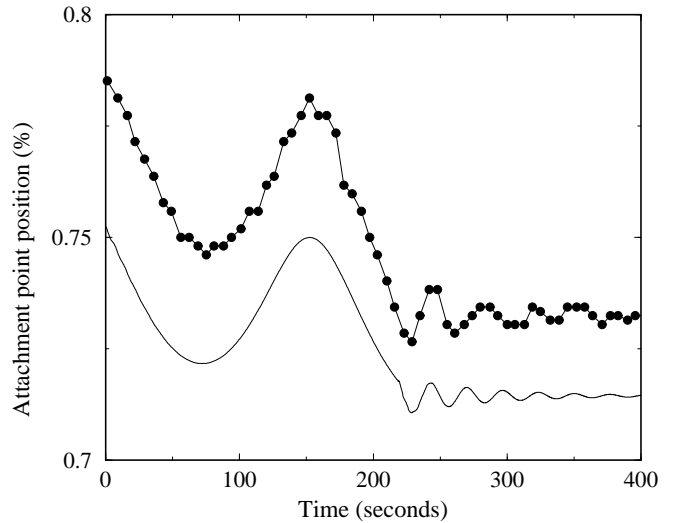}}
\caption{\small Vortex becoming pinned during its precession.  Top curve:
measurements.  Bottom curve: simulation for wire displaced 0.25 mm from
cell center, pin site at $\theta=126^\circ$, friction coefficient $\alpha=0.05$.}
\label{f:pinned}
\end{center}
\end{figure}

In Schwarz' simulations, depinning events have characteristic signatures
\cite{Schwarzheli}.  As shown in Figure \ref{f:pinned}, we find remarkably
similar experimental signals, although at the onset of pinning rather than at
depinning.  The precession oscillations at the left give way to faster, lower
amplitude oscillations upon pinning.  The energy dissipation also ceases, as
expected once the vortex stops moving.   The constant background signal that
results is convenient for detecting the pinning oscillations experimentally. 
Upon depinning, any special signature is superimposed on both the oscillations
from vortex precession and the linear change from decay of the trapped
vorticity.  In addition, if an external perturbation causes the depinning,
other irregularities may occur in the decay signal just after the event.

The fast oscillations upon pinning arise from Kelvin waves on the free vortex
segment.  Kelvin originally treated of vortex waves on an infinite straight
vortex with a sinusoidal distortion of wavelength $k$ \cite{Kelvin}.  The
distortion spirals about the core at frequency $\omega=\kappa k^2 A/4\pi$. 
Here $A \approx \ln(\frac{1}{ka})$ and $a$ is the core radius.  The boundary
conditions of our experimental setup require a standing wave, which can be
obtained by combining waves spiraling in opposite directions.  With a node on
the cell wall and maximum displacement on the wire, we require
$k=\frac{\pi}{2d}$ for the lowest mode, with $d$ the distance from the wire to
the pin site.

In Figure \ref{f:pinned}, the free vortex segment is longest at 131 seconds,
and from this position travels for 85\% of one period before pinning.  From the
amplitude of the precession oscillations, the estimated wire displacement is
0.25 mm.  Then the distance from the wire to the pin site is 1.7 mm.  The
corresponding straight vortex mode has $\omega=0.11$ rad/s and period 58
seconds, substantially longer than the observed period of 37 seconds.  The
discrepancy comes largely from the effect of the cylindrical container, which
forces the vortex to bend. The vortex line tension, or how the energy changes
with shape, depends strongly on configuration. Indeed, even in a parallel-plane
geometry, boundary effects can lead to curvature that significantly changes the
Kelvin oscillation period.  

Pinning simulations for the present experimental geometry confirm the period
change.  As described in the previous section, we begin with a steadily
precessing vortex and suddenly fix its free end.  We choose the pin location to
match the measured pin event.  We note that, in keeping with our minimum angle
simulations, the pin site in Figure \ref{f:pinned}. is near where the free
vortex is longest.  This is also true for the other long pinning events we
observe.  The simulations show oscillation of
the free portion, coupled to motion of the attachment point up and down the
wire, with period 29 seconds.  The discrepancy between the simulation and
experiment may come from our uncertain knowledge of the wire's displacement
and the pin site's location in the cell.  This is the clearest observation to
date of a Kelvin wave on a single superfluid vortex line.

The damping of the Kelvin mode is many orders of magnitude too high to come
from mutual friction with the normal fluid.  Since one end is stationary at the
wall, it is likely that the dissipation comes from the other end of the vortex
moving up and down along the wire.

\begin{figure}[b]
\begin{center}
\scalebox{.48}{\includegraphics{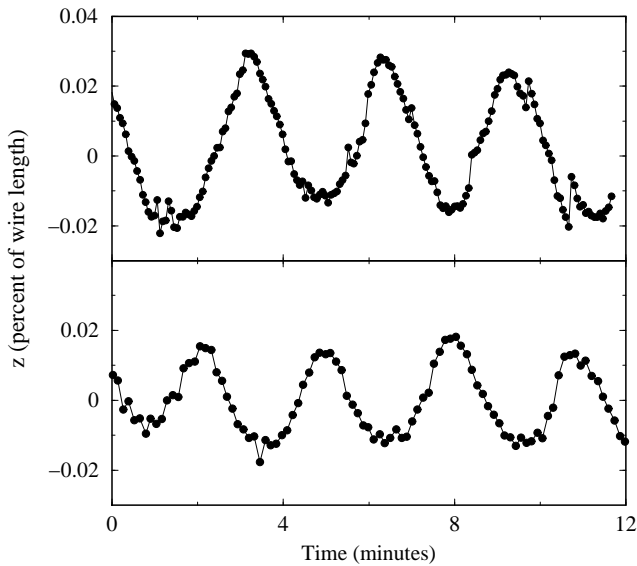}}
\caption{\small Asymmetry in the precession oscillations, for wires
A (upper) and C (lower).  Note the relative roundedness of the troughs and
sharpness of the peaks.}
\label{f:asym}
\end{center}
\end{figure}

\section{Asymmetric Oscillations}
Another feature predicted by Schwarz \cite{Schwarzheli} is a slight asymmetry
in the precession signature.  The vortex velocity depends on the local fluid
velocity.  If the wire were centered in the cylinder, the fluid velocity would
have no angular dependence.  With the off-center wire, boundary effects
increase the velocity  where the wire is closest to the wall, so the vortex
moves more quickly on this side.  In the precession signal, the fastest motion
occurs at the top of each cycle, when the free vortex is short.  The slowest
precession is at the lowest trapped circulation, when the free vortex is
longest.  The result is slightly sharpened peaks and rounded troughs.  

Figure \ref{f:asym} shows exactly this structure,
for wires A and C.   To
remove the overall slope caused by energy loss, a line has been 
subtracted from each data set.  Both the oscillation amplitude and the
asymmetry are larger for wire A than for wire C, implying that wire A is
farther off center.  Observing the asymmetry in the precession signal is
encouraging for the eventual goal of being able to track more complicated
motion of a single vortex.

\begin{figure}[bth]
\begin{center}
\psfrag{lowT}{\LARGE 340 mK\hspace{2in} 380 mK}
\psfrag{t1}{\LARGE 524 s}
\psfrag{t2}{\LARGE 556 s}
\scalebox{.5}{\includegraphics{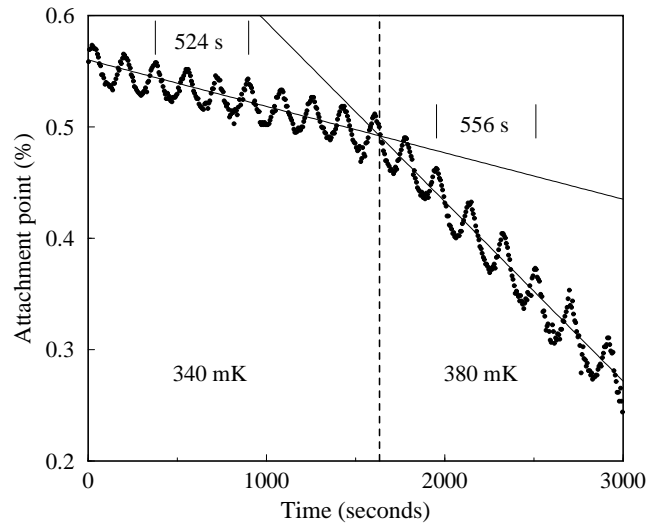}}
\caption{\small Dependence of precession rate and dissipation on temperature.
The solid lines are guides to the eye.  The times for three periods
are indicated, with the period increasing a small but noticeable amount
with temperature.}
\label{f:precdep}
\end{center}
\end{figure}

\section{Temperature dependence}
Another prediction involves temperature dependence.  In addition to its
interaction with the local velocity field, a vortex line feels a friction-like 
force proportional to its velocity.  The friction force determines the energy
loss, and hence the linear circulation decrease that appears along with vortex
precession. Experimentally the friction coefficient $\alpha$ depends strongly
on temperature.   Figure \ref{f:precdep} illustrates the dramatic change in
energy loss rate upon changing the temperature from 340 mK to 380 mK.

Computer simulations reveal that, while moving downward more rapidly, the
vortex also precesses less quickly  \cite{Schwarzheli}.  We now find
experimentally that, as temperature increases, the dissipation rate and the
precession period also increase.  While the period change is small---only
6\% in Figure \ref{f:precdep}---it is certainly outside our measurement error.

One difference between the simulations and experiments is the source of the
friction. In the simulations, the dissipation comes from mutual friction, the
interaction between the vortex and the normal fluid.  This is not the case in
experiments.  In our temperature range, mutual friction is many orders of
magnitude too small to account for the observed dissipation \cite{Reif}.  
Instead, the most likely dissipation mechanism is the rubbing of the end of the
vortex along the wall.  It is interesting that the relationship between the
precession rate and the dissipation holds nonetheless.

Schwarz finds that as dissipation increases, the precession rate slows 
\cite{Schwarzheli}. In the limiting case of large dissipation, the vortex moves
directly downward without precessing at all. Our wire B have this behavior; its
circulation decay is very fast and exhibits no oscillations.

\begin{figure}[h]
\begin{center}
\scalebox{.3}{\includegraphics{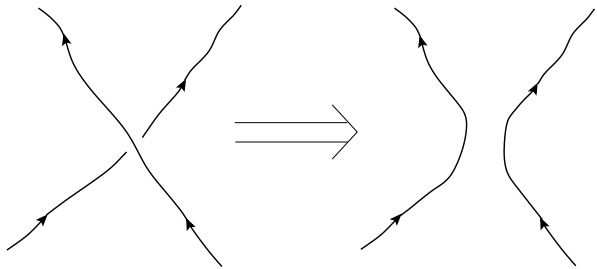}}
\caption{\small Vortices reconnecting upon collision.}
\label{f:recon}
\end{center}
\end{figure}

\section{Initiating precession}
In the work described here we usually shake the cryostat to initiate the vortex
motion.  However, in superfluid ${}^3$He \cite{3Hepaper} and occasionally in
our present measurements, a trapped vortex may leave the wire without 
deliberate external perturbation.  How the unwinding process begins remains
unknown.  Schwarz suggests that a free vortex in the cell collides with the
$n=1$ vortex along the wire \cite{Schwarzheli}.  Close approach of vortices 
often leads to a runaway attraction ending in reconnection \cite{Schwarz}, 
illustrated in Figure
\ref{f:recon}.  The vortices are cut at the crossing point, and the segments
leading in and out are reattached in the opposite way.  A reconnection
involving one free vortex and one trapped around the wire would create two
partially attached vortices.  

\begin{figure}[h]
\begin{center}
\scalebox{0.5}{\includegraphics{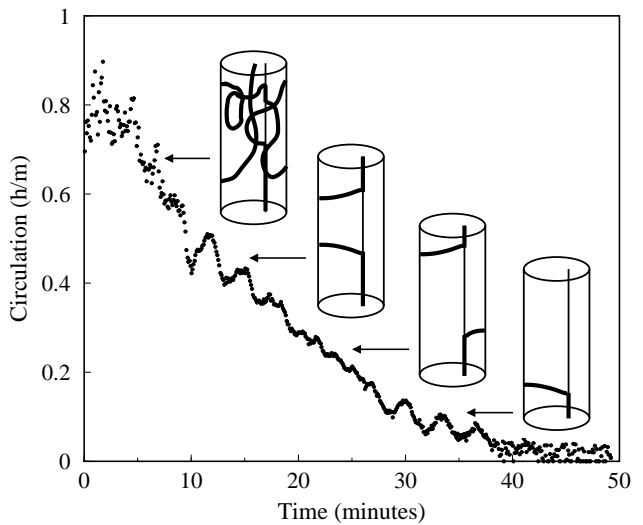}}
\caption{\small Vortices unwinding from both ends of the wire simultaneously.
The schematics illustrate possible vortex configurations at the times
indicated.}
\label{f:2partial}
\end{center}
\end{figure}

Figure \ref{f:2partial} shows what we believe is two unwinding partial
vortices.  The cryostat was rotated immediately before the data pictured. From
the irregular circulation signal for the first few minutes we hypothesize that
a jumble of vortices pervades the cell.  After 10 minutes precession begins,
with the usual period near three minutes and large amplitude oscillations. 
This is consistent with two partially attached vortices, which are on the same
side of the wire at the same time.  Additional structure appears in the decay,
and by 20 minutes the frequency has doubled, with the amplitude dropping
dramatically.  One explanation is that the two precessing vortices are now out
of phase.  If one is long while the other is short, their effects on the
trapped vortex length partly cancel, reducing the oscillation amplitude.  (The
cancellation cannot be complete.  Since the superfluid velocity field, and
hence the precession rate, depends on the position within the cell, the two
vortices cannot remain perfectly out of phase with each other.)  The frequency
doubling occurs because the two vortices attain an equivalent configuration,
with their roles transposed, after each traverses only half of the cell.
Finally, after 30 minutes, the oscillations return to the usual precession
frequency.   These are cleaner than the large oscillations earlier in the
decay, and signify that one of the vortices has completed its unwinding,
returning the cell to a single-vortex situation.

We have seen one other similar signature.  Since the vortex precession is
usually much cleaner than this, we believe that our more typical signatures
involve only one partially attached vortex.  This means that the unwinding
process generally does {\em not} start from a free vortex collision, since even
unwinding that begins without shaking rarely leads to signatures like that of
Figure \ref{f:2partial}.  We have calculated numerically the energy in the
flow field for two partially attached vortices if they are on the same or
opposite sides of the cell.  The out of phase configuration is energetically
favorable, which is consistent with its appearance in our observations.  We have
not yet investigated how the vortices change from in phase to out of phase.

\section{Conclusions}
We present data on smooth vortex precession in superfluid ${}^4$He.  A key
difference between our work and other vibrating wire measurements is that our
wires are located substantially away from the axis of the container.  We verify
several numerical predictions for precession signatures:  behavior at pinning
events, the oscillation asymmetry from the off-center wire, and the
relationship between the vertical motion of the trapped vortex and its
precession period.  The repeatable, smooth vortex precession can serve
as a probe in further experiments on dynamics of one or a few vortices.

Our measurements also address the question of how the precession begins.  We
see a distinctive signature for two partially attached vortices.  Its absence
in the bulk of our measurements means that this configuration rarely initiates
the precession.  A more common situation appears to be a partially dislodged
vortex which may return to the wire or change to terminate on the cell wall.
Since many of our vortex decays begin with literally shaking the cryostat, our
typical detachment mechanism may differ from behavior in the absence of such
strong external perturbations.  Better controlled studies of how the vortex
first comes off the wire remain to be done.


\begin{thebibliography}{99}

\bibitem{gen} H.J. Lugt, {\em Vortex Flow in Nature and Technology}
(John Wiley \& Sons, New York, 1983).

\bibitem{vortmeth1} L. Ying and P. Zhang, {\em Vortex methods} (Science
Press, Beijing, 1997).

\bibitem{vortmeth2} S.M. Belotserkovskii and I.K. Lifanov, {\em Method
of discrete vortices} (CRC Press, Boca Raton, 1993).

\bibitem{2rings} H. Yamade and T. Matsui, ``Mutual slip-through of a pair
of vortex rings," {\em Phys. Fluids} {\bf 22}, 1245 (1979).

\bibitem{gen2} S.I. Green, {\em Fluid Vortices} (Kluwer Academic Publishers,
Dordrecht, 1995), Chapter 1.

\bibitem{BEC} B.P. Anderson, P.C. Haljan, C.E. Wieman, and E.A. Cornell,
``Vortex precession in Bose-Einstein condensates:  Observations with
filled and empty cores," {\em Phys. Rev. Lett.} {\bf 85}, 2857 (2000).

\bibitem{Finnemore} M. Breitwisch and D. Finnemore, ``Pinning of a single
Abrikosov vortex in superconducting Nb thin films using artificially induced
pinning sites," {\em Phys. Rev.} {\bf B62}, 671 (2000) and references therein.

\bibitem{Vinen} W.F. Vinen, ``The detection of a single quantum of
circulation in liquid helium II," {\em Proc. Roy. Soc. London}{\bf A260},
218 (1961).

\bibitem{WZ} S.C. Whitmore and W. Zimmermann, Jr., ``Observation of quantized
circulation of superfluid helium," {\em Phys. Rev.}
{\bf 166}, 181 (1968).

\bibitem{3HeQC} J.C. Davis, J.D. Close, R. Zieve, and R.E. Packard, 
``Observation of quantized circulation in superfluid ${}^3$He-B,"
{\em Phys. Rev. Lett.} {\bf 66}, 329 (1991).

\bibitem{helicopter} R.J. Zieve et al., ``Precession of a single vortex line in
superfluid ${}^3$He," {\em Phys. Rev. Lett.} {\bf 68}, 1327 (1992).

\bibitem{3Hepaper} R.J. Zieve et al., ``Investigation of quantized circulation
in superfluid ${}^3$He-B," {\em J. Low Temp. Phys.} {\bf 91}, 315 (1993).

\bibitem{4Hepaper} R.J. Zieve, J.D. Close, J.C. Davis, and R.E. Packard, ``New
experiments on the quantum of circulation in ${}^4$He," {\em J. Low Temp.
Phys.} {\bf 90}, 243 (1993).

\bibitem{Schwarzheli} K.W. Schwarz, ``Unwinding of a single quantized vortex
from a wire," {\em Phys. Rev.} {\bf B47}, 12030 (1993).

\bibitem{Schwarz} K.W. Schwarz, ``Three-dimensional vortex dynamics in
superfluid ${}^4$He:  Line-line and line-boundary interactions," {\em Phys.
Rev.} {\bf B31}, 5782 (1985).

\bibitem{lukesim} R.J. Zieve and L.A.K. Donev, ``Stable Vortex Configurations
in a Cylinder," {\em J. Low Temp. Phys.} {\bf 121}, 199 (2000).

\bibitem{Karn} P.W. Karn, D.R. Starks, and W. Zimmermann, Jr., ``Observation
of quantization of circulation in rotating superfluid ${}^4$He," {\em Phys.
Rev.} {\bf B21}, 1797 (1980).

\bibitem{Kelvin} Thomson, W., ``Vibrations of a columnar vortex," 
{\em Phil. Mag.} {\bf 10}, 155 (1880).

\bibitem{Reif} G.W. Rayfield and F. Reif, ``Quantized vortex rings in 
superfluid helium," {\em Phys. Rev.} {\bf
136}, A1194 (1964).

\end{thebibliography}
\end{document}